\documentclass{article}

\usepackage{spconf,amsmath,multirow,booktabs}
\usepackage{graphicx}
\usepackage{comment}\title{Learning to Use Multiple Black-Box Experts for ASR Training}
\usepackage{xcolor}
\usepackage{graphicx}
\usepackage{subcaption}
\usepackage[normalem]{ulem}
\usepackage{amsfonts}

\ninept

\title{Cross-utterance ASR Rescoring with Graph-based Label Propagation}

\name{\parbox{\textwidth}{\centering
Srinath Tankasala$^{1 \dagger}$ \qquad Long Chen$^{2 \dagger}$%
\thanks{$^{\dagger}$Equal contribution. This first author was an intern at Amazon.}
\qquad Andreas Stolcke$^{2}$ \qquad Anirudh Raju$^{2}$ \qquad Qianli Deng$^{2}$ \\
Chander Chandak$^{2}$ \qquad Aparna Khare$^{2}$ \qquad Roland Maas$^{2}$ \qquad Venkatesh Ravichandran$^{2 \dagger}$}}

\address{$^1$The University of Texas at Austin, USA \hspace{10mm}
      $^2$Amazon Alexa AI, USA}

\usepackage{algorithm,algcompatible}

\begin{document}
%
\maketitle
\begin{abstract}
We propose a novel approach for ASR N-best hypothesis rescoring with graph-based label propagation by leveraging cross-utterance acoustic similarity. In contrast to conventional neural language model (LM) based ASR rescoring/reranking models, our approach focuses on acoustic information and conducts the rescoring collaboratively among utterances, instead of individually. Experiments on the VCTK dataset demonstrate that our approach consistently improves ASR performance, as well as fairness across speaker groups with different accents. Our approach provides a low-cost solution for mitigating the majoritarian bias of ASR systems, without the need to train new domain- or accent-specific models.
\end{abstract}
\section{Introduction}
\label{sec:intro}

AI virtual assistants are used widely today, allowing customers to access a large variety of services and experiences by voice. Automatic speech recognition (ASR), which converts spoken utterances into textual form, is key to enable this human-machine interaction. 

In a conventional ASR system, a two-pass system is employed where the first pass produces N-best hypotheses \cite{schwartz1991nbest}, and the second pass rescores/reranks them to produce the final ASR hypothesis. Conventionally, an end-to-end deep neural acoustic model, such as a recurrent neural network transducer (RNN-T) \cite{alex2012rnnt,alex2013rnnt}, is used in the first pass, while a language model (LM) \cite{tomas2010lm} trained on a large text dataset is employed for the rescoring stage. However, these conventional components face several challenges. First, the first-pass deep neural acoustic model is typically trained with datasets such as LibriSpeech \cite{librispeech} to optimize an average loss over all training samples, which usually introduces a majoritarian bias and leads to worse ASR performance for underrepresented groups (such as nonnative or regional accents, idiosyncratic pronunciations, or special domains). This fairness concern has been widely discussed in a variety of machine learning domains such as face recognition \cite{facefairness}, recommendation systems \cite{recommandfairness}, as well as ASR \cite{asrfairness} and speaker recognition \cite{sidfairness}. As only textual information is available to the LM, this majoritarian bias introduced due to acoustic factors, such as accents, cannot be fully addressed with LM-based rescoring. Second, the conventional rescoring system only considers a single utterance during rescoring. While some LM approaches take context into account, no acoustic information beyond the current utterance is used, thereby making it impossible to take advantage of acoustic patterns at the domain, household, or user level. 

We propose an ASR rescoring method in which multiple utterances effectively collaborate in deciding the most likely hypotheses by leveraging cross-utterance acoustic similarity. Graph-based label propagation (graph-LP) \cite{zhou2003learning} has been widely used in fields like computer vision \cite{lpcv2013,lpcv2019} and  natural language processing \cite{lpnlp2010}, and has recently been applied to speech classification tasks such as speaker identification (SID) \cite{chen2021graph,chen2022graph}. The intuition behind graph-LP applied to speech utterance classification is to exploit pairwise similarities to ensure a consistent overall labeling of utterances. In the case of SID, this can be used to extend a partial speaker labeling of utterances to an unlabeled set, based on speaker embedding similarity. Similarly, for ASR we should be able to obtain evidence about the correctness of hypotheses by comparing utterances acoustically. If two utterances sound similar then they should have similar hypotheses, and conversely, if they sound dissimilar, their hypotheses should be too. This could help especially in the case of idiosyncratic pronunciations or accents. If two utterances contain a low-frequency phrase in their hypotheses or any word with a nonstandard and therefore low-scoring pronunciation, and they share an acoustically similar segment, then the correctness of those hypotheses is mutually consistent and therefore more likely. However, unlike for standard classification problems, directly applying graph-LP to the ASR task is nontrivial, as the label space for ASR is infinite, consisting of all strings over the vocabulary. 

To make the problem tractable, we limit ourselves to a finite set of labels, i.e., N-best hypotheses for each utterance, and take their union across utterances as the label set. We create graphs with utterances as the nodes, and utterance-utterance similarities as the edge weights. We introduce a distance metric based on dynamic time warping (DTW) \cite{shokoohi2017generalizing} to measure the utterance-utterance similarity, and apply graph-LP to predict the overall best hypotheses. We demonstrate that this approach can improve the ASR model performance and fairness, without tuning embedding or training any domain- or accent-specific adapted models. To the best of our knowledge, this is the first work utilizing utterance-utterance acoustic similarity to carry out cross-utterance ASR hypothesis rescoring. 

In contrast to other recent work that considers the cross-utterance information for ASR rescoring \cite{chiu2021crossutterance,pmlr-v119-huang20k,zweig2009new}, which utilizes context/semantic information and assumes utterances to be from the same dialog, our approach utilizes acoustic information and can be applied to utterances from disparate contexts. There has been prior work on ASR rescoring that uses acoustic information \cite{chan2016rescore,ankur2020rescore,sainath2019rescore}, but these approaches deal with utterances individually. In contrast, our approach focuses on utterance-utterance acoustic similarity and uses it for joint rescoring. Our method is not replacing the existing LM-based rescoring systems (which can be applied prior to cross-utterance rescoring), but provides an alternative and low-cost solution for leveraging non-local acoustic information in rescoring. Our method would most naturally be employed in offline processing of speech utterance collections, e.g., for teacher label creation in semi-supervised learning.

\section{Proposed Method}
\label{sec:format}

\subsection{Problem setup}
\label{ssec:setup}
In large speech datasets, including those from AI virtual assistants, it is common for groups of utterances to have some or all of their words in common; we call these overlapped utterances and transcripts, respectively. Our goal is to take advantage of this overlap in joint rescoring of such utterance sets, using graph-LP. 
Given a dataset with multiple groups of overlapped utterances, we build a graph for each group, with utterances ($u_1, u_2, \ldots, u_M$) corresponding to the graph nodes.  For a tractable graph-LP solution, we need a label set for the graph nodes that is finite. Therefore, we use an existing ASR model to generate $N$-best hypotheses, giving us a total of $M\cdot N$ hypotheses. Since the utterances in the graph are similar, these hypotheses may be redundant. We create a hypothesis index set $\mathcal{H} = \{1,2,3, \ldots, C\}$ referring to the unique hypotheses across the $M$ utterances. Each utterance $u_i$ will have a label vector $y_i \in R^C$, indicating the likelihood of the possible hypotheses.  The label set $\hat{Y} = \{y_1, \ldots, y_M\} \subset R^C$ is initialized based on the ASR model's predicted confidence in each hypothesis. Let $X = \{x_1, \ldots, x_M\}$ be the acoustic embeddings of the utterances and $\hat{Y}^{(0)} = \{y^0_1, \ldots, y^0_M\}$ be the initial labels of the utterances. The goal is to improve (rescore) $\hat{Y}$ based on $\hat{Y}^{(0)}$ and $X$.

\subsection{Utterance-utterance distance modeling}
\label{ssec:utt_utt_distance}

The utterance-utterance distance metric is the key for graph-LP \cite{chen2021graph,chen2022graph}. Our goal is to improve the performance for any given ASR system without tuning or retraining the embeddings. We employ an RNN-T model to generate both utterance embeddings and hypotheses. Frame-wise outputs from the RNN-T encoder are used as the utterance embeddings. In order to model the aggregated distance over all frames, we compute the distance between two sets of frame-level embeddings, $x_i \in R^{T_1\times D}$ and $x_j \in R^{T_2\times D}$, by using a dependent dynamic time warping (d-DTW) distance \cite{shokoohi2017generalizing} with length normalization: 
\begin{align}
d{\text -}DTW_{norm}(x_i, x_j) = \frac{\min_{(p,q)\in P} \sqrt{\sum_{(p,q)} d(x_{ip}, x_{jq})^2}}{\text{max}(len(x_i),len(x_j))} \label{eq:dDTW_normalized}
\end{align}
where, $(p,q),\ p\in[1,T_1],\ q\in[1,T_2]$, is the warping path that matches time indices in $x_i$ to time indices in $x_j$. $d(x_{ip}, x_{jq})$ is the frame-wise distance function between the $D$-dimensional vectors $x_{ip}$ and $x_{jq}$; we use Euclidean distance in our experiments. $\text{max}(len(x_i),len(x_j))$ is the length normalization term, with $len(\cdot)$ giving the number of frames in an utterance. We also tested other embeddings and metrics, such as traditional DTW distance (with and without length normalization). However, the metric defined above was found to be suited best as our graph edge function, as it had a high correlation with the Levenshtein distance between the corresponding utterance transcripts, as discussed in Section \ref{ssec:eer_dtw_study}.

\subsection{Graph construction}
\label{ssec:graph_construction}
We create a fully connected graph for each group of utterances with similar audio transcripts. For each graph, a graph node represents an utterance, and an edge connecting two nodes represents the acoustic distance of the corresponding utterances, using d-DTW. 
In development, we tried applying a soft radial basis function kernel to the distances as the edge weight function, similar to \cite{chen2021graph}.
However, we found that binarizing the edge weights to 0/1 values gave better results.
Specifically, we threshold the distances between utterances. The final affinity matrix $W$ with edge weights between nodes $i,j$, is defined as:
\begin{align}
    W_{ij} = \begin{cases}
    1 \quad \text{if}\ d{\text -}DTW_{norm}(x_i, x_j) < \Theta \\
    0 \quad \text{if}\ d{\text -}DTW_{norm}(x_i, x_j) \geq \Theta
    \end{cases} \label{eq:Wij}
\end{align}
where $d{\text -}DTW_{norm}(x_i, x_j)$ is the normalized dependent DTW distance and $\Theta$ is the threshold to determine if two utterances are close enough in the embedding space. We optimize $\Theta$ on a development set.

\subsection{Label propagation}
\label{ssec:label_propagation}

Label propagation (LP) is a transductive graph-based semi-supervised learning (graph-SSL) approach where labels are propagated from ``labeled'' nodes to unlabeled nodes. LP tries to find a joint labelling $\hat{Y}^*$ for all graph nodes such that (a) $\hat{Y}^*$ is close to $\hat{Y}^{(0)}$; and (b) the labels are smooth over the graph, i.e., they do not differ drastically between neighbours. This is typically done by optimizing the following objective function:
\begin{align}
    \hat{Y}^* = \underset{f}{\text{argmin}} \ ||f-Y||_2^2 + \lambda \cdot trace(f^T L_{sym} f) \label{eq:lpopt}
\end{align}
where $Y$ is the input of known labels, $f$ is the labeling solution and $\lambda$ is a regularization hyperparameter. $L_{sym}$ is the symmetric normalized Laplacian graph matrix, i.e., $L_{sym} = \mathcal{I} - \Delta^{-1/2}W\Delta^{-1/2}$, where $\Delta$ is the degree diagonal matrix with $\Delta_{ii} = \sum_{j=1}^M W_{ij}$. To solve Equation~\eqref{eq:lpopt}, an iterative algorithm by Zhou et al.\ \cite{zhou2003learning} is used, as follows: 
\begin{algorithm}
    \caption{Label propagation}
    \begin{algorithmic} [1]
        \STATE Compute the affinity matrix $W$ if $i \neq j$ \& $W_{ii}=0$; 
        \STATE Compute matrix $S = \Delta^{-1/2}W\Delta^{-1/2}$
        \STATE Initialize $\hat{Y}^{(0)}$ with each row $(\hat{Y}^{(0)})_i = y_i$, where $y_i$ is a soft label vector for utterance $i$ (see Section \ref{ssec:applying_lp})
        \STATE Iterate $\hat{Y}^{(t+1)} = \alpha S \hat{Y}^{(t)} + (1-\alpha)\hat{Y}^{(0)}$ until convergence, where $\alpha\in (0,1)$
        \STATE Label each point $u_i$ with $y_i = \underset{j\leq C}{argmax}\ \hat{Y}^{(\infty)}_{ij}$
    \end{algorithmic}
\end{algorithm}
\vspace{-2mm}

\subsection{Graph-LP for cross-utterance ASR rescoring}
\label{ssec:applying_lp}

Graph-LP relies on an initial label matrix $\hat{Y}^{(0)}$. Typically in graph-SSL work \cite{chen2021graph, chen2022graph}, ground truth or ``labeled'' samples are included in the graph with hard (i.e., one-hot) initialized labels, to serve as the ``seeds'' for propagating information to unlabeled samples. In our scenario, there is no ground truth. Instead, we initialize the label vector for all utterances with soft labels over the hypothesis set $\mathcal{H}$. To do this we use the log likelihood scores of hypotheses as computed by the RNN-T model. Assume for a given utterance $u_i$ the model predicts the hypotheses $\{h_1, \ldots, h_B\}$ with log likelihoods $\{s_1, \ldots, s_B\}$, where $B$ is the beam size ($B\geq N$). For each hypothesis $k$, we compute the score $p_k$ as
\begin{align}
    p_k = \mathrm{softmax}(s_k) = \frac{e^{s_j}}{\sum_{k=1}^B e^{s_k}}
\end{align}
These probabilities $p_k$ corresponding to the top $N$ hypotheses are used as the soft labels $y_i \in R^C$ for utterance $u_i$, such that $y_i>0, ||y_i||_1\leq 1$. We generate $\hat{Y}^{(0)}$ by computing $y_i$ for all utterances $u_i$ in the graph. Algorithm~1 in Section~\ref{ssec:label_propagation} is then applied with $\hat{Y}^{(0)}$ as initialization.

\section{Experiments}
\label{sec:expts}

\subsection{Datasets}
\label{ssec:dataset}
We use the LibriSpeech \cite{librispeech} training dataset to train the ASR RNN-T model for embedding and hypothesis generation. Evaluation is based on the VCTK \cite{vctk2019} dataset. We further divide the VCTK utterances into development and test sets with a ratio of 1:2. The development set is used for metric and hyperparameter selection, while the test set is used for reporting ASR performance. LibriSpeech is commonly used for ASR tasks in the literature, with the majority of the speech coming from American English speakers reading audio books. The VCTK dataset is a popular dataset for accent studies, with English sentences sourced from newspapers read-out by speakers from 13 English-speaking regions. We chose these two datasets since they are mismatched in both domain and accents. We did not tune or adapt the ASR model to the VCTK data, to evaluate the efficacy of our proposed approach in improving the ASR model trained on out-of-domain data.


\subsection{Baseline and embedding generation model}
\label{ssec:baseline}
The baseline RNN-T ASR model uses a six-layer LSTM encoder with a hidden dimension of 1024, and a transcription network with two 1024-dimensional LSTM layers. We use a sentence-piece model \cite{SentencePiece2018} to generate output targets for the ASR model. The model was trained on the LibriSpeech dataset and has a word error rate (WER) of 6.05\% and 15.43\% on LibriSpeech-Clean and LibriSpeech-Other test sets, respectively. We evaluate the model on the VCTK dataset and use that as the baseline for comparing with the proposed graph-LP method, using both WER and sentence error rate (SER). Additionally, we focus on overall model performance as well as performance on different accent groups to test whether the proposed method can improve model performance and fairness.

\label{ssec:embedding_prep}
The baseline RNN-T model is also used to generate the inputs for the graph-LP algorithm.  The embeddings computed by the final RNN-T encoder layer are used for utterance-utterance distance calculation, as described in Section~\ref{ssec:utt_utt_distance}.

\subsection{Metric selection for utterance-utterance distance}
\label{ssec:eer_dtw_study}
A good utterance-utterance distance function used for graph-LP needs to satisfy the following property:
\textit{For any pair of utterances $i,j$ in the graph, the distance in the embedding space should reflect the distant between the corresponding ground-truth transcripts, e.g., embedding distance should be highly correlated with the Levenshtein distance between transcripts.}

The above property ensures that the distance function serves the ASR task, rather than measuring similarity along other dimensions, such as speaker ID or acoustic environment. To quantify this property, we borrow the concept of equal error rate (EER) used for metric learning and verification tasks \cite{EER}. We create trials of utterance pairs from the development set with 10,000 positive and 50,000 negative pairs, where positive/negative pairs correspond to utterances having the same/different ground-truth transcripts. The utterance-utterance distance is calculated for each pair. We then find the threshold at which false accept rate (FAR) and false reject rate (FRR) are equalized, giving us EER = FAR = FRR. We also use t-SNE plots to visualize utterance similarities.

We consider two groups of candidate methods for the utterance-utterance distance function, as well as variants with and without length normalization:
\begin{itemize}
    \item Euclidean distance between the last frame embeddings emitted by the RNN-T audio encoder
    \item Traditional DTW or dependent DTW ($d{\text -}DTW$) distance between RNN-T audio encoder embeddings of all frames
\end{itemize}
The rationale behind this choice is as follows: (1) for RNN-based models, the last frame embedding from the output layer in principle could encapsulate the information of the whole audio; (2) the DTW-based distance function evaluates the time-warped distance between a given pair of sequences, intuitively reflecting the accumulated distance over all frames; (3) length normalization allows more consistent distance thresholding across different utterance lengths.  

Using the above candidate distance functions, we compute the EER for all the methods. The EER value is used to select the utterance-utterance distance metrics. The evaluation results for the distance functions described above are reported in Section~\ref{ssec:eer_results}.

\subsection{Graph-LP experiments}
\label{ssec:lp_expts}
As described in Section~\ref{ssec:graph_construction}, we aim to construct graphs by pooling utterances with similar transcripts. However, given that ASR is the task, we do not have prior access to the ground truth transcripts for the test utterances. Instead, we pool the utterances based on their baseline ASR hypotheses. To make the label propagation method scalable, we only group utterances with similar hypotheses into one graph. First, the tf-idf embeddings of all utterances are generated using the ASR 1-best hypotheses. We then use the DBSCAN algorithm \cite{dbscan1996} to identify utterance clusters and build a graph from all the utterances in one cluster. Ideally, we want the sizes of generated clusters to be within a suitable range. Too many utterances would result in large graphs with many nodes having low hypothesis overlap. If the cluster size is too small, there may be not be enough information added by considering multiple utterances in joint rescoring. We tuned the parameters of the DBSCAN algorithm on the development set to maximize the number of clusters with sizes in the range 4 to 800. Utterances that cannot be clustered are not included in graph-LP and their hypotheses are left unchanged.

Given an utterance cluster, we select the top $N=3$ hypotheses for each utterance to construct the label set $\mathcal{H}$. The label confidences are calculated using the method described in Section~\ref{ssec:applying_lp} to generate the initial labelings $\hat{Y}^{(0)}$. In the graph, we want label information to flow strongly between similar utterances. Hence, we calculate $W_{ij}$ using Equation~\eqref{eq:Wij}. To reduce computation further, we remove connections between nodes where the minimum word edit distance between the top 3 hypotheses is $>4$. Graph-LP is then applied to generate the final labels for all the nodes. We allow label sharing, i.e., the final label for an utterance can be outside its initial $N$-best hypothesis set, potentially improving results.
\section{Results}
\label{sec:results}

\subsection{Baseline model results}
\label{ssec:subhead}
Performance of the baseline RNN-T model on the VCTK dataset is shown in Table~\ref{tab: vctk_accent_eval}. We show word error rates by speaker accents. Here, WER-5best is the oracle WER of the 5-best hypotheses. We can see a significant difference between performance on American/Canadian compared to English, Scottish, and other regional accents, attributable to the LibriSpeech training dataset consisting mainly of American English speech. There is also a significant performance gap between the 1-best WER and 5-best WER ($13.98\% \rightarrow 7.89\%$), showing the potential for improvement with hypothesis rescoring.
\vspace{-2mm}


\begin{table}[tb]
    \centering
    \caption{Baseline RNN-T WER results on the VCTK dataset.}
    \vskip -1em
    \resizebox{\linewidth}{!}{
    \begin{tabular}{l r r r r r}
          \hline
          Accent & \# Speakers & \# Utterances & WER-1best & WER-5best \\
          \hline
          English & 33 & 27207 & 15.22 & 8.84 \\
          Scottish & 19 & 15184 & 16.59 & 10.18 \\
          American & 21 & 16760 & 9.41 & 4.35 \\
          Irish & 9 & 7230 & 15.48 & 8.80 \\
          Canadian & 8 & 6286 & 9.10 & 4.19 \\
          Northern Irish & 6 & 5148 & 15.41 & 8.38 \\
          South African & 4 & 3366 & 11.82 & 5.95 \\
          Indian & 3 & 2322 & 18.51 & 12.44 \\
          Others & 5 & 3643 & 16.21 & 9.17 \\
          \hline
          Overall & 108 & 87146 & 13.98 & 7.89 \\
          \hline
    \end{tabular}}
    \label{tab: vctk_accent_eval}
\end{table}

\subsection{EER and metric selection results}
\label{ssec:eer_results}

\begin{figure}[tb]
\begin{subfigure}{0.49\linewidth}
  \centering
  \includegraphics[width=0.99\linewidth]{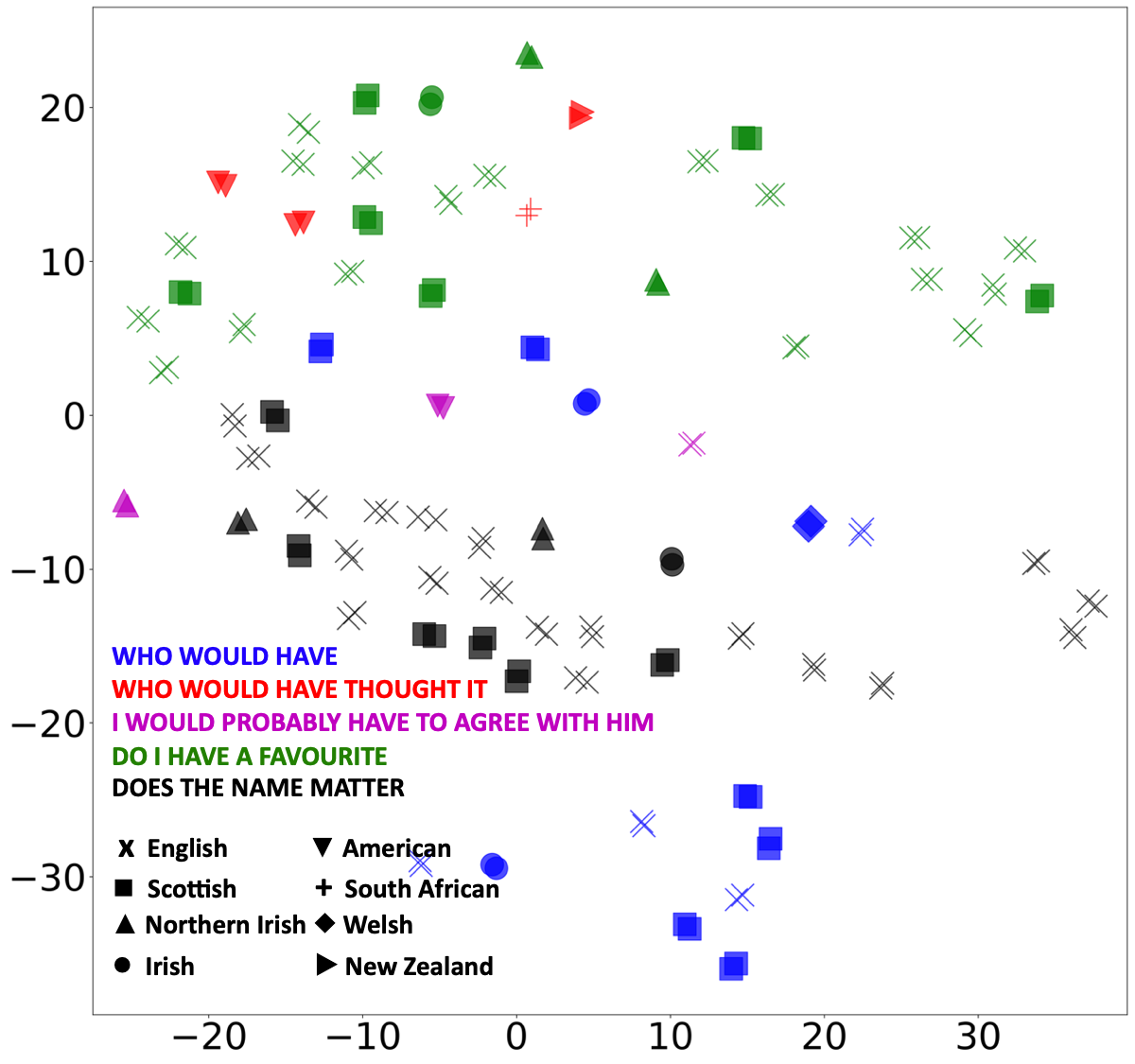}
  \vspace{-5pt}
  \caption{Last frame embedding distance}
  \label{fig:sub-first}
\end{subfigure}
\begin{subfigure}{0.49\linewidth}
  \centering
  \includegraphics[width=0.99\linewidth]{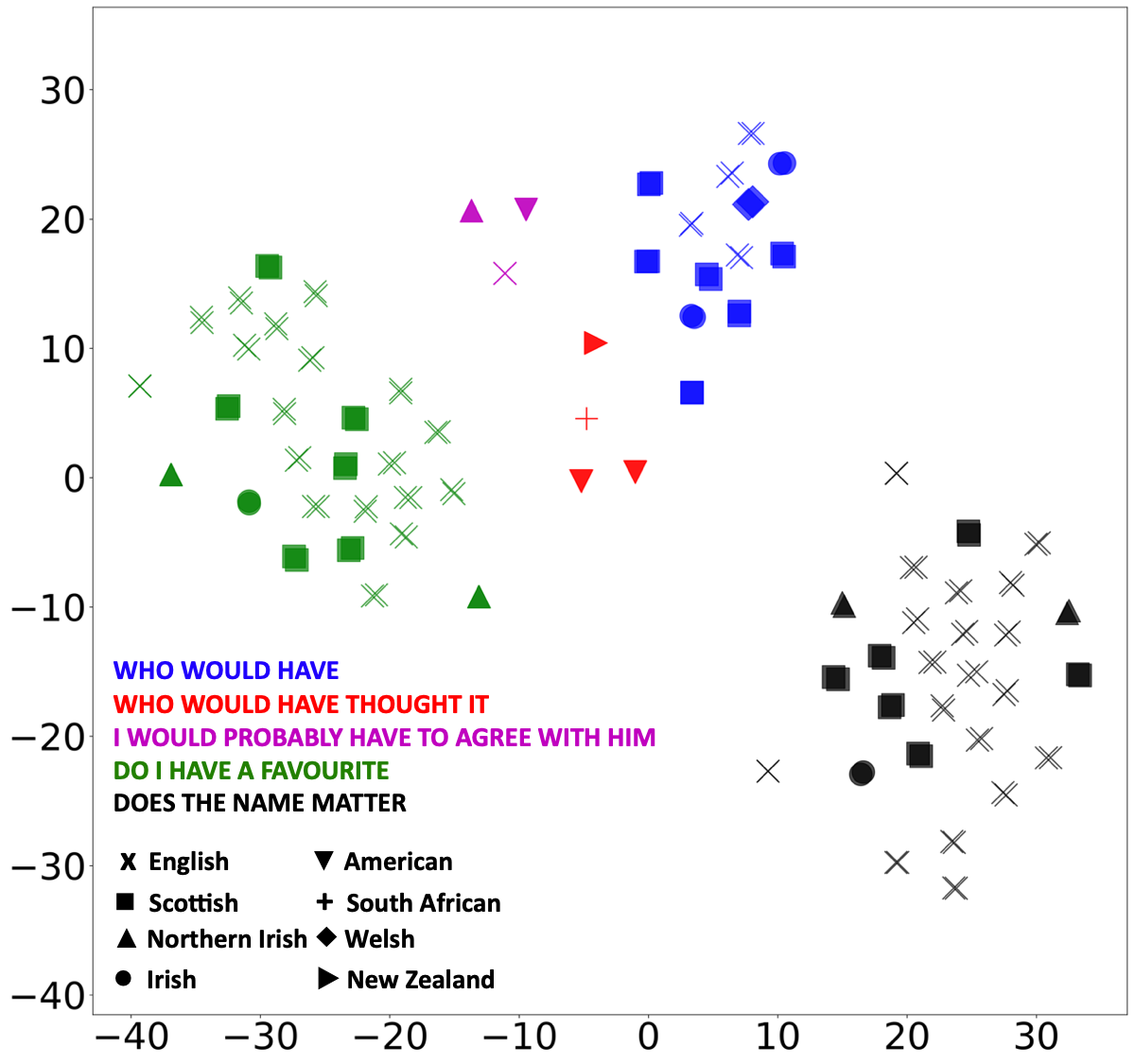} 
  \vspace{-5pt}
  \caption{All frame embedding d-DTW}
  \label{fig:sub-second}
\end{subfigure}
\vspace{-5pt}
\caption{t-SNE visualization of utterance-utterance distances. Dots represent utterances in embedding space, with color and shape coding the transcript and accent of an utterance, respectively. (a) Euclidean distance based on last-frame embeddings. (b) d-DTW distance based on all-frames embeddings. }
\label{fig:tsne}
\end{figure}

Figure~\ref{fig:tsne} visualizes a sample of VCTK utterances using t-SNE, based on various distance metrics. It clearly shows the clustering of similar utterances in the label space when using d-DTW distance based on all frame embeddings. We also observe that clusters that have more transcript overlap are closer together (e.g., blue versus red samples). This is not the case when using a distances based only on the last-frame RNN-T encoder embeddings, for which no clustering of utterances with identical audio transcripts is observed. From the visualization, we infer that the last-frame embedding distance is an unsuitable metric for constructing our graphs. Table~\ref{tab:EER} shows the EERs for same-ground-truth classification with several alternative distance metrics, with and without length normalization.  Length-normalized d-DTW achieves the lowest EER;  it is used in all graph-LP results reported here. The devtest-optimized distance threshold in Equation~\eqref{eq:Wij} is $\Theta = 1.5$, leaving about 47\% of edges remaining. 

\begin{table}[t!]
    \centering
    \caption{EERs (\%) of various acoustic utterance distance metrics without and with length normalization. LFE: Euclidean distance of last frame embeddings; DTW: traditional dynamic time warping distance; d-DTW: dependent DTW distance.}
    \vskip -1em
    \begin{tabular}{l c c c}
          \hline
          Metric & LFE & DTW & d-DTW \\
          \hline
          without length normalization & 38.78 & 17.40 & 7.48\\
          with length normalization & 36.38 & 6.34 & 4.50 \\
          \hline
    \end{tabular}
    \label{tab:EER}
\end{table}


\vspace{-2mm}
\subsection{Graph-LP rescoring results}
\label{ssec:WER_2}

\begin{table}[tb]
    \centering
    \caption{Baseline and graph-LP results based on hypothesis tf-idf clustering. WER and SER are in \%.
    The last row includes test utterances that were not included in any clusters and graph-LP.}
    \vskip -1em
    \resizebox{\linewidth}{!}{
    \begin{tabular}{l r r r r r r}
    \hline
           & \multicolumn{1}{c}{\#}  & \multicolumn{1}{c}{\#} & \multicolumn{2}{c}{Baseline} & \multicolumn{2}{c}{Graph-LP} \\
          Cluster size ($n$) & Clusters & Utterances & WER & SER & WER & SER\\
          \hline
          $n\leq$5 & 1782 & 7112 & 8.54 & 36.38 & 6.24 &26.73 \\
          $5< n \leq$10 & 1352 & 10008 & 10.26 & 40.49 & 6.77 & 27.17 \\
          $10< n \leq$50 & 837 & 14191 & 10.41 & 42.15 & 5.27 & 21.82 \\
          $n >$50 & 37 &4722 &10.16 &55.78 &4.50 &28.80 \\
          \hline
          All clustered & 4008 & 36033 &9.99 &42.33 &5.64 &25.19 \\
          All utterances & - & 58098 &13.97 &50.31 &11.14 &39.67 \\
          \hline
    \end{tabular}}
    \label{tab: LP_cluster}
\end{table}

\begin{table}[tb]
\caption{Effect of label (hypothesis) sharing on graph-LP results.}
\label{tab:hyp-sharing-results}
    \vskip -1em
\centering
\begin{tabular}{lrrrr}
\hline
Utterance set   & \multicolumn{2}{c}{Without sharing} & \multicolumn{2}{c}{With sharing} \\
& WER & SER & WER & SER \\
\hline
All clustered   & 8.75  & 35.36 & 5.64 & 25.19 \\
All utterances  & 13.17 & 45.98 & 11.14 & 39.67 \\
\hline
\end{tabular}
\end{table}


Table~\ref{tab: LP_cluster} shows results for graph-LP-based cross-utterance rescoring as described in Section~\ref{ssec:lp_expts}. We observe significant improvements in WER across cluster sizes, with an overall improvement of 43.5\% for WER and 40.5\% for SER, respectively. Clusters of larger size seem to show a bigger performance gain. This is consistent with the notion that the more related utterances are involved in graph-LP, the more additional information can be aggregated, compared to single-utterance recognition. Moreover, as shown in Table~\ref{tab:hyp-sharing-results}, sharing labels (hypotheses) among all utterances in the same cluster gives a substantial benefit, reducing WER by 35.5\% and SER by 28.8\%. Label sharing effectively recovers plausible hypotheses left out of the original N-best lists, and graph-LP allows evaluating them even though they do not have a likelihood based on the first-pass ASR.

Table~\ref{tab: LP_accent} shows ASR results for different accent groups.
(Note that Tables~\ref{tab: LP_cluster} and~\ref{tab: LP_accent} are based on the test set only, rather than all of VCTK as in Table~\ref{tab: vctk_accent_eval}.)
WER and SER across all accent groups are improved. Moreover, accent groups other than American/Canadian show larger improvements, leading to a much smaller gap between the high and low performance groups. These results demonstrate that the proposed approach is effective at mitigating the majoritarian bias of the original ASR system, improving both accuracy and fairness. 

\begin{table}[t!]
    \centering
    \caption{Baseline and graph-LP results by regional accents. WER and SER are in \%.}
    \vskip -1em
    \resizebox{\linewidth}{!}{
    \begin{tabular}{l r r r r r}
    \hline
        \multicolumn{2}{c}{} & \multicolumn{2}{c}{Baseline} & \multicolumn{2}{c}{Graph-LP} \\
          Accent & \# Utterances & WER & SER & WER & SER\\
          \hline
          English & 12960 & 10.84 & 45.58 & 5.82 & 25.83 \\
          Scottish & 7174 & 11.78 & 48.69 & 5.76 & 26.33 \\
          American & 5807 & 6.67 & 30.15 & 5.02 & 23.23 \\
          Irish & 2899 & 10.46 & 45.50 & 5.68 & 26.04 \\
          Canadian & 2151 & 6.77 & 30.96 & 5.01 & 22.55 \\
          Northern Irish & 1657 & 9.94 & 40.13 & 5.82 & 22.75 \\
          South African & 1164 & 7.82 & 34.62 & 4.68 & 21.82 \\
          Indian & 937 & 13.26 & 48.88 & 7.94 & 31.06 \\
          Others & 1284 & 10.43 & 46.11 & 5.99 & 25.62 \\
          \hline
          Overall & 36033 & 9.99 & 42.33 & 5.64 & 25.19 \\
          \hline
    \end{tabular}}
    \label{tab: LP_accent}
\end{table}

\vspace{-0.5em}

\section{Conclusions}
\label{sec:majhead}
We have proposed a cross-utterance ASR hypothesis rescoring approach based on graph-based label propagation (graph-LP). Our approach improves ASR performance by leveraging (1) cross-utterance information, especially acoustic similarity, modeled by a DTW-based distance metric and (2) joint cross-utterance rescoring enabled by graph-LP and a shared hypothesis set among utterances. The approach is designed to help ASR systems adapt to idiosyncratic pronunciations, accents, or out-of-domain content. Experiments on the VCTK dataset demonstrate that the proposed approach consistently improves overall error rates, as well as for speaker groups with specific accents. Our method
is well-suited to offline ASR settings, without requiring adaptation or fine-tuning of the baseline model.

\clearpage

\bibliographystyle{IEEEbib}

\begin{thebibliography}{10}

\bibitem{schwartz1991nbest}
Richard Schwartz and Steve Austin,
\newblock ``A comparison of several approximate algorithms for finding multiple
  (n-best) sentence hypotheses,''
\newblock in {\em Proc.\ IEEE ICASSP}, 1991, pp. 701--704.

\bibitem{alex2012rnnt}
Alex Graves,
\newblock ``Sequence transduction with recurrent neural networks,''
\newblock in {\em Proc.\ ICML}, 2012.

\bibitem{alex2013rnnt}
Alex Graves, Abdel-rahman Mohamed, and Geoffrey Hinton,
\newblock ``Speech recognition with deep recurrent neural networks,''
\newblock in {\em Proc.\ IEEE ICASSP}, 2013, pp. 6645--6649.

\bibitem{tomas2010lm}
Tomas Mikolov, Martin Karafi{\'a}t, Luk{\'a}{\v s}s Burget, Jan~Honza {\v
  C}ernock{\'y}, and Sanjeev Khudanpur,
\newblock ``Recurrent neural network based language model,''
\newblock in {\em Proc.\ Interspeech}, 2010, pp. 1045--1048.

\bibitem{librispeech}
Vassil Panayotov, Guoguo Chen, Daniel Povey, and Sanjeev Khudanpur,
\newblock ``{LibriSpeech}: An {ASR} corpus based on public domain audio
  books,''
\newblock in {\em Proc.\ IEEE ICASSP}, 2015, pp. 5206--5210.

\bibitem{facefairness}
Mei Wang, Weihong Deng, Jiani Hu, Xunqiang Tao, and Yaohai Huang,
\newblock ``Racial faces in the wild: Reducing racial bias by information
  maximization adaptation network,''
\newblock in {\em Proc.\ ICCV}, 2019, pp. 692--702.

\bibitem{recommandfairness}
Alex Beutel, Jilin Chen, Tulsee Doshi, Hai Qian, Li~Wei, Yi~Wu, Lukasz Heldt,
  Zhe Zhao, Lichan Hong, Ed~H. Chi, and Cristos Goodrow,
\newblock ``Fairness in recommendation ranking through pairwise comparisons,''
\newblock in {\em Proc.\ ACM SIGKDD Int. Conf. Knowl. Discov. Data Min.}, 2019,
  pp. 2212--2220.

\bibitem{asrfairness}
Pranav Dheram, Murugesan Ramakrishnan, Anirudh Raju, I-Fan Chen, Brian King,
  Katherine Powell, Melissa Saboowala, Karan Shetty, and Andreas Stolcke,
\newblock ``Toward fairness in speech recognition: Discovery and mitigation of
  performance disparities,''
\newblock in {\em Proc.\ Interspeech}, 2022, pp. 1268--1272.

\bibitem{sidfairness}
Hua Shen, Yuguang Yang, Guoli Sun, Ryan Langman, Eunjung Han, Jasha Droppo, and
  Andreas Stolcke,
\newblock ``Improving fairness in speaker verification via group-adapted fusion
  network,''
\newblock in {\em Proc.\ IEEE ICASSP}, 2022, pp. 7707--7081.

\bibitem{zhou2003learning}
Dengyong Zhou, Olivier Bousquet, Thomas Lal, Jason Weston, and Bernhard
  Sch{\"o}lkopf,
\newblock ``Learning with local and global consistency,''
\newblock in {\em Proc.\ NIPS}, 2003, pp. 321--328.

\bibitem{lpcv2013}
Yan Wang, Rongrong Ji, and Shih-Fu Chang,
\newblock ``Label propagation from {ImageNet} to {3D} point clouds,''
\newblock in {\em Proc.\ IEEE Conference on Computer Vision and Pattern
  Recognition}, 2013, pp. 3135--3142.

\bibitem{lpcv2019}
Bin Liu, Zhirong Wu, Han Hu, and Stephen Lin,
\newblock ``Deep metric transfer for label propagation with limited annotated
  data,''
\newblock in {\em Proc.\ IEEE/CVF International Conference on Computer Vision
  Workshop (ICCVW)}, 2019, pp. 1317--1326.

\bibitem{lpnlp2010}
Amarnag Subramanya, Slav Petrov, and Fernando Pereira,
\newblock ``Efficient graph-based semi-supervised learning of structured
  tagging models,''
\newblock in {\em Proc.\ EMNLP}, 2010, p. 167–176.

\bibitem{chen2021graph}
Long Chen, Venkatesh Ravichandran, and Andreas Stolcke,
\newblock ``Graph-based label propagation for semi-supervised speaker
  identification,''
\newblock in {\em Proc.\ Interspeech}, 2021, pp. 4588--4592.

\bibitem{chen2022graph}
Long Chen, Yixiong Meng, Venkatesh Ravichandran, and Andreas Stolcke,
\newblock ``Graph-based multi-view fusion and local adaptation: Mitigating
  within-household confusability for speaker identification,''
\newblock in {\em Proc.\ Interspeech}, 2022, pp. 4805--4809.

\bibitem{shokoohi2017generalizing}
Mohammad Shokoohi-Yekta, Bing Hu, Hongxia Jin, Jun Wang, and Eamonn Keogh,
\newblock ``Generalizing {DTW} to the multi-dimensional case requires an
  adaptive approach,''
\newblock {\em Data mining and knowledge discovery}, vol. 31, no. 1, pp. 1--31,
  2017.

\bibitem{chiu2021crossutterance}
Shih-Hsuan Chiu, Tien-Hong Lo, Fu-An Chao, and Berlin Chen,
\newblock ``Cross-utterance reranking models with {BERT} and graph
  convolutional networks for conversational speech recognition,''
\newblock in {\em Proc.\ Asia-Pacific Signal and Information Processing
  Association Annual Summit and Conference}, 2021, pp. 1104--1110,
\newblock also arXiv:2106.06922.

\bibitem{pmlr-v119-huang20k}
Hengguan Huang, Fuzhao Xue, Hao Wang, and Ye~Wang,
\newblock ``Deep graph random process for relational-thinking-based speech
  recognition,''
\newblock in {\em Proc. \ ICML}, 2020, vol. 119, pp. 4531--4541.

\bibitem{zweig2009new}
Geoffrey Zweig,
\newblock ``New methods for the analysis of repeated utterances,''
\newblock in {\em Proc.\ Interspeech}, 2009, pp. 2791--2794.

\bibitem{chan2016rescore}
William Chan, Navdeep Jaitly, Quoc Le, and Oriol Vinyals,
\newblock ``Listen, {Attend} and {Spell}: A neural network for large vocabulary
  conversational speech recognition,''
\newblock in {\em Proc.\ IEEE ICASSP}, 2016, pp. 4960--4964.

\bibitem{ankur2020rescore}
Ankur Gandhe and Ariya Rastrow,
\newblock ``Audio-attention discriminative language model for {ASR}
  rescoring,''
\newblock in {\em Proc.\ IEEE ICASSP}, 2020, pp. 7944--7948.

\bibitem{sainath2019rescore}
Tara~N. Sainath, Ruoming Pang, David Rybach, Yanzhang He, Rohit Prabhavalkar,
  Wei Li, Mirk{\'o} Visonta, Qiao Liang, Trevor Strohman, and Yonghui Wu,
\newblock ``Two-pass end-to-end speech recognition,''
\newblock in {\em Proc.\ Interspeech}, 2019, pp. 2773--2777.

\bibitem{vctk2019}
Junichi Yamagishi, Christophe Veaux, and Kirsten MacDonald,
\newblock ``{CSTR VCTK Corpus}: English multi-speaker corpus for {CSTR} {Voice
  Cloning Toolkit} (version 0.92),''
\newblock {\em University of Edinburgh. The Centre for Speech Technology
  Research (CSTR)}, 2019.

\bibitem{SentencePiece2018}
Taku Kudo and John Richardson,
\newblock ``{SentencePiece}: {A} simple and language independent subword
  tokenizer and detokenizer for neural text processing,''
\newblock in {\em Proc.\ EMNLP}, 2018, pp. 66--71.

\bibitem{EER}
N.~Br{\"u}ummer and J.~{du Preez},
\newblock ``Application-independent evaluation of speaker detection,''
\newblock {\em Computer Speech and Language}, vol. 20, pp. 230–275, 2006.

\bibitem{dbscan1996}
Martin Ester, Hans-Peter Kriegel, J\"{o}rg Sander, and Xiaowei Xu,
\newblock ``A density-based algorithm for discovering clusters in large spatial
  databases with noise,''
\newblock in {\em Proc.\ KDD}, 1996, p. 226–231.

\end{thebibliography}

\end{document}